%% file: paper.tex
\documentclass[journal]{IEEEtran}
\usepackage[pdftex]{graphicx}
\usepackage[cmex10]{amsmath}
\usepackage{amssymb}
\usepackage{bm}
\usepackage{cite}
\usepackage{hyperref}
\usepackage{textcomp}
\usepackage[squaren]{SIunits}
\usepackage{todonotes}
\usepackage{multirow}
\usepackage[acronym]{glossaries} 
\usepackage{dsfont}
\makeglossaries 
\glsdisablehyper

\usepackage{xcolor}
\usepackage{color}
\definecolor{light}{rgb}{.1,.6,.1}

\def\deg{{^\circ}}
\def\vec#1{\ensuremath{\bm{{#1}}}}
\def\mat#1{\vec{#1}}
\def\mat#1{\vec{#1}}
\DeclareMathOperator*{\argmax}{\arg\!\max}

\newglossaryentry{ibm}{name={IBM},description={ideal binary mask},
first={\glsentrydesc{ibm} (\glsentrytext{ibm})},
plural={IBMs},
descriptionplural={ideal binary masks},
firstplural={\glsentrydescplural{ibm} (\glsentryplural{ibm})}}
\newglossaryentry{itd}{name={ITD},description={interaural time difference},
first={\glsentrydesc{itd} (\glsentrytext{itd})},
plural={ITDs},
descriptionplural={interaural time differences},
firstplural={\glsentrydescplural{itd} (\glsentryplural{itd})}}
\newglossaryentry{ild}{name={ILD},description={interaural level difference},
first={\glsentrydesc{ild} (\glsentrytext{ild})},
plural={ILDs},
descriptionplural={interaural level differences},
firstplural={\glsentrydescplural{ild} (\glsentryplural{ild})}}
\newacronym{2d}{2D}{two-dimensional}
\newacronym{tf}{T-F}{time-frequency}
\newacronym{snr}{SNR}{signal-to-noise ratio}
\newacronym{ams}{AMS}{amplitude modulation spectrogram}
\newacronym{asa}{ASA}{auditory scene analysis}
\newacronym{erb}{ERB}{equivalent rectangular bandwidth}
\newacronym{ccf}{CCF}{cross-correlation function}
\newacronym{casa}{CASA}{computational auditory scene analysis}
\newglossaryentry{prior}{name=\emph{a priori},description={}}
\newglossaryentry{post}{name=\emph{a posteriori},description={}}
\newglossaryentry{hfa}{name=HIT\,-\,FA,description={}}
\newacronym{dnn}{DNN}{deep neural network}
\newacronym{rasta}{RASTA-PLP}{relative spectral transform and perceptual linear prediction}
\newacronym{mfcc}{MFCCs}{mel-frequency cepstral coefficients}
\newacronym{spp}{SPP}{speech presence probability}
\newacronym{tpp}{TPP}{target presence probability}
\newacronym{drr}{DRR}{direct-to-reverberant ratio}
\newacronym{hats}{HATS}{head and torso simulator}
\newglossaryentry{brir}{name={BRIR},description={binaural room impulse response},
first={\glsentrydesc{brir} (\glsentrytext{brir})},
plural={BRIRs},
descriptionplural={binaural room impulse responses},
firstplural={\glsentrydescplural{brir} (\glsentryplural{brir})}}
\newglossaryentry{hrtf}{name={HRTF},description={head related transfer function},
first={\glsentrydesc{hrtf} (\glsentrytext{hrtf})},
plural={HRTFs},
descriptionplural={head related transfer functions},
firstplural={\glsentrydescplural{hrtf} (\glsentryplural{hrtf})}}
\newglossaryentry{hrir}{name={HRIR},description={head related impulse response},
first={\glsentrydesc{hrir} (\glsentrytext{hrir})},
plural={HRIRs},
descriptionplural={head related impulse responses},
firstplural={\glsentrydescplural{hrir} (\glsentryplural{hrir})}}
\newacronym{psd}{PSD}{power spectral density}
\newacronym{gmm}{GMM}{Gaussian mixture model}
\newacronym{kemar}{KEMAR}{Knowles Electronic Manikin for Acoustic Research}
\newacronym{lc}{LC}{local criterion}
\newacronym{rms}{RMS}{root mean square}
\newacronym{rmse}{RMSE}{root mean square error}
\newacronym{pdf}{PDF}{probability density function}
\newacronym{dft}{DFT}{discrete Fourier transform}
\newacronym{stft}{STFT}{short-time discrete Fourier transform}
\newacronym{em}{EM}{expectation-maximization}
\newacronym{mct}{MCT}{multi-conditional training}
\newacronym{ml}{ML}{most likely}
\newacronym{maa}{MAA}{minimum audible angle}
\newacronym{ubm}{UBM}{universal background model}

\begin{document}
%
\title{Robust Binaural Localization of a Target Sound Source by Combining Spectral Source Models and Deep Neural Networks}
%
%
%
\author{Ning Ma,
            Jose A. Gonzalez
            and Guy J. Brown
\thanks{This work was supported by the EC FP7 project {\sc Two!Ears} under grant agreement No.~618075. N. Ma and G. J. Brown are with the Department of Computer Science, University of Sheffield, Sheffield, UK, S1 4DP (email: \{n.ma, g.j.brown\}@sheffield.ac.uk). J. A. Gonzalez is with the Department of Languages and Computer Sciences, University of Malaga, Campus de Teatinos, Malaga 29071, Spain (email: j.gonzalez@uma.es). This work was carried out while Gonzalez was working at the University of Sheffield.}
}

\markboth{IEEE/ACM Transactions on Audio, Speech and Language Processing}%
{Ma \MakeLowercase{\textit{et al.}}: Robust binaural localisation of a target sound source by combining spectral source models and deep neural networks}




\maketitle



\begin{abstract}
Despite there being clear evidence for top-down (e.g., attentional) effects in biological spatial hearing, relatively few machine hearing systems exploit top-down model-based knowledge in sound localisation. This paper addresses this issue by proposing a novel framework for binaural sound localisation that combines model-based information about the spectral characteristics of sound sources and deep neural networks (DNNs). A target source model and a background source model are first estimated during a training phase using spectral features extracted from sound signals in isolation. When the identity of the background source is not available, a universal background model can be used. During testing, the source models are used jointly to explain the mixed observations and improve the localisation process by selectively weighting source azimuth posteriors output by a DNN-based localisation system. To address the possible mismatch between training and testing, a model adaptation process is further employed on-the-fly during testing, which adapts the background model parameters directly from the noisy observations in an iterative manner. The proposed system therefore combines model-based and data-driven information flow within a single computational framework. The evaluation task involved localisation of a target speech source in the presence of an interfering source and room reverberation. Our experiments show that by exploiting model-based information in this way, sound localisation performance can be improved substantially under various noisy and reverberant conditions.
\end{abstract}


\IEEEpeerreviewmaketitle

\input{01_Introduction}

\input{02_System}

\input{03_Localisation}

\input{04_Evaluation}

\input{05_Results}

\input{06_Conclusion}


\ifCLASSOPTIONcaptionsoff
  \newpage
\fi



%

\bibliographystyle{IEEEtran}
\bibliography{journal_abrv_short,twoears}

%

\begin{IEEEbiography}[{\includegraphics[width=1in,height=1.25in,clip,keepaspectratio]{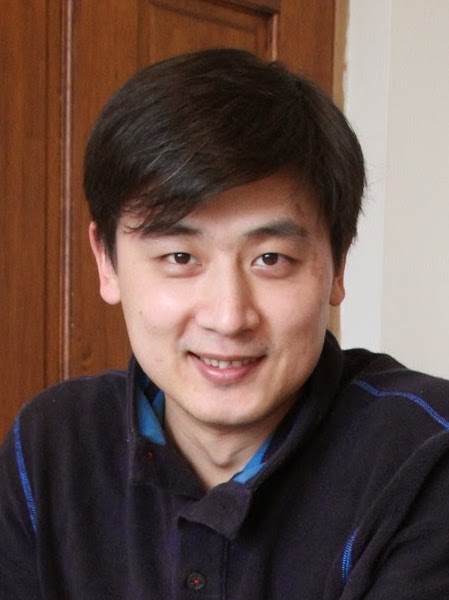}}]{Ning Ma}
obtained a MSc (distinction) in Advanced Computer Science in 2003 and a PhD in hearing-inspired approaches to automatic speech recognition in 2008, both from the University of Sheffield. He has been a visiting research scientist at the University of Washington, Seattle, and a Research Fellow at the MRC Institute of Hearing Research, working on auditory scene analysis with cochlear implants. Since 2015 he is a Research Fellow at the University of Sheffield, working on computational hearing. His research interests include robust automatic speech recognition, computational auditory scene analysis, and hearing impairment. He has authored or coauthored over 40 papers in these areas.
\end{IEEEbiography}


\begin{IEEEbiography}[{\includegraphics[width=1in,height=1.25in,clip,keepaspectratio]{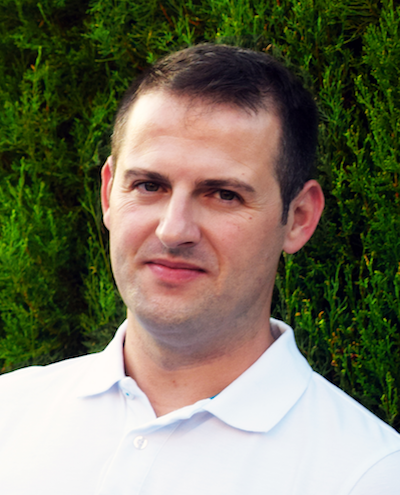}}]{Jose A. Gonzalez}
received the B.Sc. and Ph.D. degrees in Computer Science, both from the University of Granada, Spain, in 2006 and 2013, respectively.  During his Ph.D. he did two research stays at the Speech and Hearing Research Group, University of Sheffield, U.K., studying missing data approaches for noise robust speech recognition. From 2013 to 2017 he was a Research Associate at the University of Sheffield working on silent speech technology with special focus on direct speech synthesis from speech-related biosignals. Since October 2017, he is a lecturer at the Department of Languages and Computer Science, University of Malaga, Spain.
\end{IEEEbiography}

\begin{IEEEbiography}[{\includegraphics[width=1in,height=1.25in,clip,keepaspectratio]{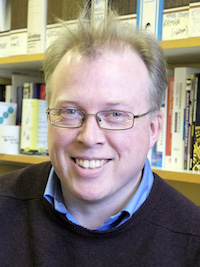}}]{Guy J. Brown}
obtained a BSc (Hons) Applied Science from Sheffield City Polytechnic in 1984 and a PhD in Computer Science from the University of Sheffield in 1992. He was appointed to a Chair in the Department of Computer Science, University of Sheffield, in 2013. He has held visiting appointments at LIMSI-CNRS (France), Ohio State University (USA), Helsinki University of Technology (Finland) and ATR (Japan). His research interests include computational auditory scene analysis, speech perception, hearing impairment and acoustic monitoring for medical applications. He has authored more than 100 papers and is the co-editor (with Prof. DeLiang Wang) of the IEEE book ``Computational auditory scene analysis: principles, algorithms and applications''.
\end{IEEEbiography}




\end{document}

%% file: 01_Introduction.tex
\section{Introduction}
\label{s:intro}



\IEEEPARstart{I}{t} has been long established that human listeners use both bottom-up (data-driven) and top-down (model-based) information in order to understand an acoustic scene (e.g., \cite{Bre1990}). More specifically, both kinds of information are needed in order to answer `what' and `where' questions about the acoustic scene, i.e. \textit{what} the identities of the sound sources are, and \textit{where} they are in space. Many machine hearing studies have proposed computational approaches for answering these questions, by combining techniques for source separation, classification and sound localisation \cite{WangBrown2006}. However, in such machine systems, data-driven and model-based mechanisms are typically much less well-integrated than they appear to be in biological hearing. The current paper addresses this issue, by proposing a novel machine hearing system that tightly integrates knowledge about source spectral characteristics with a mechanism for binaural localisation. The resulting system improves binaural sound localisation under challenging acoustic conditions in which multiple sound sources and room reverberation are present. 



Psychophysical studies of human hearing have found evidence for top-down effects in sound localisation. For example, listeners take less time to localise a target sound when it is preceded by a cue on the same side of the head, suggesting that sound localisation is enhanced by covert orienting \cite{spence1994}. More recently, based on a review of psychophysical data, Bronkhorst \cite{Bronkhorst2015} proposed a conceptual model of hearing in which top-down models inform the selection of binaural cues needed for localisation of specific sounds. Physiological studies in animals have also shown that sound localisation is modulated by top-down mechanisms. Studies of the barn owl have shown that selective attention influences sound localisation, including orienting behaviour such as body and head movements. More specifically, neural responses in the midbrain of the owl are enhanced if they are associated with the location of behaviourally relevant stimuli, such as a source of food \cite{gaese2002}. Likewise, neural circuitry for gaze control can exert a top-down effect on the responses of auditory neurons that are turned to particular spatial locations \cite{Winkowski:2006rm}. In summary, these psychological and physiological findings suggest that mechanisms of spatial hearing are tightly integrated with top-down, cross-modal processing in the brain. 



In contrast, relatively few systems for machine hearing exploit top-down model-based information in source localisation. In~\cite{MaEtAl2017dnn}, Ma et al. proposed a robust binaural localisation system based on \gls{dnn} for localisation of multiple sources. However, no source models were used and the system reported azimuth estimates for all directional sources. In \cite{Chr2009}, Christensen et al. exploit a pitch cue to identify local spectro-temporal fragments that are considered to be dominated by a single source, before integrating binaural cues over such fragments. The integrated binaural cues were more robust for localising multiple speakers, but no explicit model-based information was exploited. Mandel et al.~\cite{MandelEtAl2010} have proposed a probabilistic model for joint sound source localisation and separation based on interaural spatial cues and binary masking. Given the number of sources, the system iteratively updates \glspl{gmm} of spatial cues using the \gls{em} algorithm. However, the benefit of the system was demonstrated in terms of source separation rather than localisation. Two systems for binaural localisation of multiple sources recently proposed by \cite{WoodruffWang2012,WoodruffWang2013} use statistical frameworks to jointly perform localisation and pitch-based segregation. However, they do not use information about source characteristics other than through statistical models of pitch dynamics and binaural cues. In \cite{Bar2005, Ma2013}, model-based information is explicitly employed together with binaural cues, but the task in those studies was automatic speech recognition rather than sound localisation. Closest to the approach proposed here is the attention-driven model of sound localisation proposed by \cite{liu2011}, in which top-down connections from a cortical model are used to potentiate responses to attended locations. However, attentional control in their model is driven by a simple neural circuit that fixates on sounds arriving from the same spatial location, and their approach is currently unable to localise multiple sound sources.

The current paper proposes a framework for sound localisation in which model-based information about the spectral characteristics of sound sources in the acoustic scene is used to selectively weight binaural cues. Models for the target and the background sources are first estimated in a training stage, using spectral features computed from source signals in isolation. When the identity of the background source is not available, a universal background model can be used. The source models are then used during the testing stage to jointly explain the mixed observations in terms of which spectral features belong to each source (target or masker), and thereby improve the localisation process by selectively weighting binaural cues in each time-frequency bin. In our previous work \cite{MaEtAl2015topdown} model-based information was incorporated in a GMM-based localisation system, whereas in this study it was used in a state-of-the-art DNN-based system~\cite{MaEtAl2017dnn}.

The proposed system also addressed the possible mismatch due to power level differences between training and testing. An iterative adaptation process is employed on-the-fly to estimate a frequency-dependent scaling factor for the background model parameters directly from the mixed signals. The proposed system therefore combines data-driven and model-based information flow within a single computational framework. We show that by exploiting source models in this way, sound localisation performance can be improved under conditions in which multiple sources and room reverberation are present.

The rest of the paper is organised as follows. In the next section we give a system overview and describe features used for source localisation and for modelling source spectral characteristics. A source localisation framework is then presented in Section~\ref{s:loc} which allows model-based information of source spectral characteristics to be incorporated. Evaluation and experiments are described in Section~\ref{s:eval}. Section~\ref{s:results} discusses the results in detail. Finally, Section~\ref{s:conc} concludes the paper and makes suggestions for future work.

%% file: 02_System.tex

\section{System overview}
\label{s:system}

\begin{figure}
\centerline{\includegraphics[width=\columnwidth]{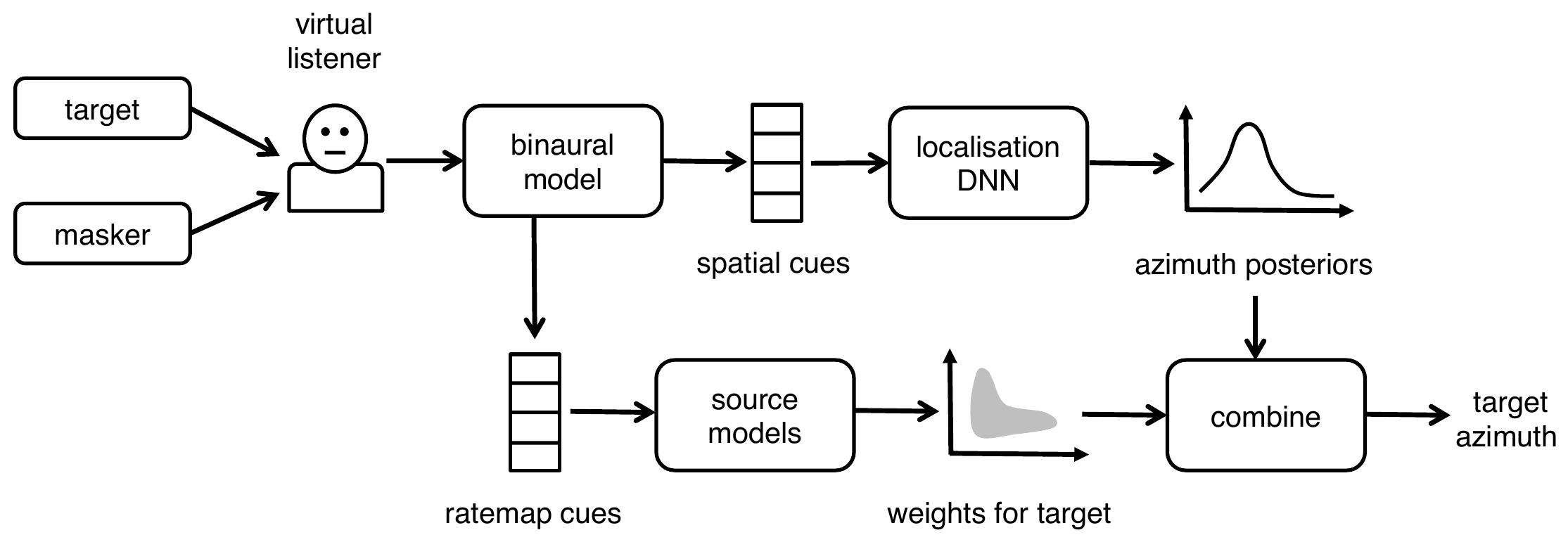}}
\caption{Schematic diagram of the proposed system.}
\label{f:system}
\end{figure}

Figure~\ref{f:system} shows a schematic diagram of the proposed binaural sound localisation system. A target source and an interfering source (masker) are spatialised in a virtual environment, which allows sounds to be placed in the full 360$\,\deg$ azimuth range around a virtual listener. In this study, the prior knowledge of the target source is assumed to be known so that the system can ``attend'' to the target for localisation. No prior knowledge of the masker is assumed when a \gls{ubm} is used, but the knowledge of the masker can be exploited by the framework when available (see Section \ref{ss:source_model} for more details).

The binaural ear signals are analysed by a binaural auditory model, which consists of a bank of 32 overlapping Gammatone filters with centre frequencies uniformly spaced on the \gls{erb} scale between 80\,Hz and 8\,kHz~\cite{WangBrown2006}. Inner hair cell function was approximated by half-wave rectification. Subsequently, a cross-correlation between the right and left ears was computed independently for each of the 32 frequency bands using overlapping frames of 20\,ms with a shift of 10\,ms. The \gls{ccf} was further normalised by the autocorrelation value at lag zero as described in \cite{MayvandeParKohlrausch11a} and evaluated for time lags in the range of $\pm$1.1\,ms.

Two binaural features, \glspl{itd} and \glspl{ild}, are typically used in binaural localisation systems \cite{Blauert97}. \gls{itd} is estimated as the lag corresponding to the maximum in the \gls{ccf}. However, it has been shown that the \gls{ccf} outperforms the \glspl{itd} in state-of-the-art \gls{dnn}-based localisation systems \cite{MaEtAl2015dnn,MaEtAl2017dnn}. Thus, in this work we use the entire \gls{ccf} as localisation features. For signals sampled at a rate of 16\,kHz, the \gls{ccf} with a lag range of $\pm1$\,ms produced a 33-dimensional binaural feature space for each frequency band. This was supplemented by the \gls{ild}, which corresponds to the energy ratio between the left and right ears within each analysis window, expressed in $\deci\bel$.  The final 34-dimensional (34D) feature vectors were passed to a \gls{dnn}-based localisation system which estimates posterior probabilities for each azimuth in the full 360$\,\deg$ azimuth range. \glspl{dnn} were trained using sound mixtures consisting of the target and masker sources rendered in a virtual acoustic environment. 

Ratemap features~\cite{Bro1994} were used to model source spectral characteristics. A ratemap is a spectro-temporal representation of auditory nerve firing rate, extracted from the inner hair cell output of each frequency channel by leaky integration and downsampling (see Fig.~\ref{f:noise} for examples). For the binaural signals used here, the ratemap features were computed for each ear channel separately and then averaged across the two ears. They were finally log-compressed to form 32D feature vectors. The source ratemap features were exploited together with a target source model and a background source model to estimate a set of localisation weights, in order to bias the system towards localising the target source. In the next section we describe the localisation framework in detail.

%% file: 03_Localisation.tex

\section{Localisation with top-down source models}
\label{s:loc}

In the following we use $P$ for probability, $p$ for density functions and $C$ for cumulative distribution functions.

\subsection{Localisation model}

\glspl{dnn} were used to model statistical relationships between the binaural features and corresponding azimuth angles~\cite{MaEtAl2015dnn,MaEtAl2017dnn}. Unlike many other binaural sound localisation systems (e.g. \cite{WoodruffWang12,MayvandeParKohlrausch13}), this study does not assume that sound sources are located only in the frontal-hemifield. Instead it considers 72 azimuth angles $\phi$ in the full $360\deg$ azimuth range ($5\deg$ steps). 

A separate DNN was trained for each of the 32 frequency bands, largely because this study is concerned with localisation of multiple sources. Although simultaneous sources overlap in time, within a local time frame each frequency band is mostly dominated by a single source\footnote{See Bregman's notion of `exclusive allocation'~\cite{Bre1990}. Note in reverberant environments this is only a crude approximation}. By adopting a separate DNN for each frequency band, it is possible to train the DNNs using single-source data without having to construct multi-source training data.

The DNN model was largely based on a state-of-art binaural localisation system described in \cite{MaEtAl2017dnn}. In brief, each \gls{dnn} consists of an input layer, two hidden layers, and an output layer. The input layer contained 34 nodes and each node was assumed to be a Gaussian random variable with zero mean and unit variance. The hidden layers had sigmoid activation functions with 128 hidden nodes. The output layer contained 72 nodes corresponding to the 72 azimuth angles in the full 360$\,\deg$ azimuth range, with a 5$\,\deg$ step. A `softmax' activation function was applied at the output layer. 



Given the observed localisation feature vector $\vec{o}_{tf}$ (the 34D localisation features described in Section~\ref{s:system}) at time frame $t$ and frequency band $f$, the posterior probability of azimuth angle $P(\phi | \vec{o}_{tf})$ is estimated using the \gls{dnn} trained for each frequency band $f$. The posteriors are then integrated across frequency to produce the probability of azimuth $\phi$ given features $\vec{o}_t= [\vec{o}_{t1}^\top, \dots, \vec{o}_{t32}^\top]^\top$ of the entire frequency range at time $t$,
\begin{equation}
P(\phi|\vec{o}_t) =  \frac{\prod\nolimits_f  P(\phi|\vec{o}_{tf}) ^ {\omega_{tf}} }{ P\left(\vec{o}_t\right)}
\label{e:post-angle}
\end{equation}
where 
\begin{equation}
P(\vec{o}_t) = \sum\nolimits_{\phi}\prod\nolimits_f  P(\phi|\vec{o}_{tf}) ^ {\omega_{tf}} 
\label{e:prob-obs}
\end{equation}

Here $\omega_{tf} \in [0,1]$ is introduced for selectively weighting the contribution of binaural cues from each time-frequency bin in order to localise the attended target source in the presence of competing sources. When $\omega_{tf}$ is $0$ the time-frequency bin will be excluded from localisation of the target source. The next section will discuss in detail how {model-based knowledge about sound sources} can be used to jointly estimate the weighting factors.

Assuming that the target sound source is stationary, the frame posteriors are further averaged across time to produce a posterior distribution $P(\phi)$ of sound source activity given a segment of signal consisting of $T$ time frames
\begin{equation}
P(\phi | \vec{o}_{1\dots T}) = \frac{1}{T} \sum_t^{t+T-1} P(\phi|\vec{o}_{t})
\label{e:post}
\end{equation}
The target location is considered to be the azimuth $\hat{\phi}$ that maximises Eq.~(\ref{e:post})
\begin{equation}
\hat{\phi} = \argmax_\phi P(\phi | \vec{o}_{1\dots T})
\end{equation}

\subsection{Exploiting model-based information}
\label{ss:source_model}

In the presence of multiple sources, the binaural cues computed from spectro-temporal regions dominated by the  masking sources will bias the localisation decision towards the location of the maskers, thus leading to localisation errors. To address this issue, we propose an `attentional' mechanism that exploits prior information about the spectral characteristics of the acoustic sources in order to weight the binaural cues towards localising the target source. First, the prior source models are used to estimate the probability of each \gls{tf} bin of the observed signal being dominated by the energy of the target source. These probabilities are then employed during sound localisation to selectively weight the binaural cues: cues extracted from \gls{tf} regions dominated by the masking sources are penalised during localisation.

Let $\vec{y}_t= [y_{t1}, \dots, y_{tD}]$ denote the ratemap feature vector (see Section II) extracted from the observed signals at frame $t$, where $D$ is the feature dimension. For simplification we will omit the dependence on the time index $t$ for the remainder of this section. Similarly, let $\vec{x}$ and $\vec{n}$ denote the ratemap feature vectors of the underlying target and masker, respectively.  In the log-ratemap domain, the relationship between these variables can be approximated as 
\begin{equation}
y_{f} \approx \max(x_{f}, n_{f})
\label{e:log-max}
\end{equation}
where $f$ is the frequency band within $[1 \dots D]$. This is known as the \emph{log-max} model~\cite{Var1990,Ren2010}, which is an approximation of the exact interaction model between two sound sources when they are expressed in a log-compressed spectral domain. According to this model, the effect of the masker sound on the target can be modelled as a kind of spectral masking.

Let $\vec{\omega}$ denote the localisation weights to be estimated, and each of its elements $\omega_{f} \in [0,1]$ indicates whether $y_{f}$ is dominated by the energy of the target source $x_{f}$ ($\omega_{f} \approx 1$) or the masker source $n_{f}$ ($\omega_{f} \approx 0$). From a probabilistic point of view, and under the restrictions imposed by the \emph{log-max} model, $\omega_{f}$ corresponds to the following \emph{a posteriori} probability
\begin{equation}
\omega_{f} \triangleq P(x_{f} \ge n_{f}|\vec{y}) \equiv P(x_{f}=y_{f}, n_{f} \le y_{f}|\vec{y})
\label{e:weights}
\end{equation}
To estimate this probability, prior models for the spectral characteristics of the sound sources are used. Each sound source $s=1, \ldots, \mathcal{S}$ is modelled using a \gls{gmm} $\lambda_s$ with $K_s$ mixtures. Because the identity of the target source is known, the probability of the observation given the target source \gls{gmm} $\lambda_x$ is computed simply as
\begin{equation}
p(\vec{y}|\lambda_x)= \sum_k^{K_x} P(k|\lambda_x) \mathcal{N}\left(\vec{y}; \vec{\mu}_x^{(k)}, \mat{\Sigma}_x^{(k)}\right)
\label{e:gmm-labmbdax}
\end{equation}
where $\vec{\mu}_x^{(k)}$ and $\mat{\Sigma}_x^{(k)}$ are the mean and covariance of the $k^{th}$ component for the target source \gls{gmm}.

For the case of the masker source we will distinguish two cases. First, if there is only one masker source in the acoustic scene and its identity is known \textit{a priori}, $p(\vec{y}|\lambda_n)$ can be computed as in (\ref{e:gmm-labmbdax}) but using the \gls{gmm} for the known masker, $\lambda_n$. If the identity of the masker is unknown, we estimate a \gls{ubm} $\lambda_{ubm}$ using signals from various sources (see Section \ref{ss:source_model_training} for more details about the estimation of the UBM).

Using $p(\vec{y}|\lambda_x)$ and $p(\vec{y}|\lambda_n)$, Eq.~(\ref{e:weights}) for computing the localisation weights $\omega_{f}$ can be rewritten as follows (see \cite{Reddy2004,Reddy2007,Gonzalez2017}):
\begin{equation}
\omega_{f} = \sum_{k_x} \sum_{k_n} \gamma^{(k_x,k_n)} P(x_{f}=y_{f}, n_{f} \le y_{f}|k_x,k_n, \vec{y})
\label{e:mask-double-sum}
\end{equation}
Here, $k_x$ and $k_n$ denote the index for the mixture components in $\lambda_x$ and $\lambda_n$, $P(x_{f}=y_{f}, n_{f} \le y_{f}|k_x,k_n, \vec{y})$ is the \gls{tpp} and $\gamma^{(k_x,k_n)}$ is defined as the posterior probability 
\begin{equation}
\gamma^{(k_x,k_n)} \triangleq P(k_x,k_y|\vec{y})= \frac{p(\vec{y}|k_x,k_n) P(k_x) P(k_n)}{\sum_{k_x',k_n'} p(\vec{y}|k_x',k_n') P(k_x') P(k_n')}
\label{e:mask-post-prob}
\end{equation}
Here we have omitted the dependence on the models $\lambda_x$ and $\lambda_n$ in order to simplify the notation.

Assuming the frequency bands are conditionally independent given the mixture components, $p(\vec{y}|k_x,k_n)$ can be expressed as
\begin{equation}
p(\vec{y}|k_x,k_n) = \prod_{f=1}^D p(y_{f}|k_x,k_n)
\end{equation}
with $p(y_{f}|k_x,k_n)$ being the following marginal distribution
\begin{align}
p(y_{f}|k_x,k_n)  =  \iint p(y_{f}|x_{f},n_{f}) p(x_{f}|k_x) p(n_{f}|k_n) dx_{f} dn_{f}
\label{e:obs-prob-ytf}
\end{align}
The terms $p(x_{f}|k_x)$ and $p(n_{f}|k_n)$ of the equation can be directly calculated using the \glspl{gmm} $\lambda_x$ and $\lambda_n$. Using the \emph{log-max} model, $p(y_{f}|x_{f},n_{f})$ can be expressed as
\begin{align}
p(y_{f}|x_{f},n_{f}) &= \delta(y_{f} - \max(x_{f},n_{f})) \nonumber \\
&= \delta(y_{f} - x_{f}) \mathds{1}_{n_{f} \leq x_{f}} + \delta(y_{f} - n_{f}) \mathds{1}_{x_{f} < n_{f}}
\label{e:post_y_xn}
\end{align}
where $\delta(\cdot)$ is the Dirac delta function and $\mathds{1}_\mathcal{C}$ is an indicator function which equals 1 when the condition $\mathcal{C}$ is true and 0 otherwise. As shown in \cite{Gonzalez2017}, (\ref{e:obs-prob-ytf}) then becomes
\begin{align}
p(y_{f}|k_x,k_n)  =  p_x(y_{f}|k_x)C_n(y_{f}|k_n) + p_n(y_{f}|k_n)C_x(y_{f}|k_x)
\label{e:prob_obs_y}
\end{align}
where $p_x$ and $p_n$ are the Gaussian probability functions and $C_x$ and $C_n$ are the corresponding cumulative distribution functions.

Using Bayes' rule, the \gls{tpp} in (\ref{e:mask-double-sum}) can be written as
\begin{align}
& P(x_{f}=y_{f}, n_{f} \le y_{f}|k_x,k_n, \vec{y}) \nonumber \\ 
& \, = \frac{p(x_{f}=y_{f}, n_{f} \le y_{f}|k_x,k_n)}{p(x_{f}=y_{f}, n_{f} \le y_{f}|k_x,k_n)+p(n_{f}=y_{f}, x_{f} < y_{f}|k_x,k_n)} \nonumber \\
& \, = \frac{p_x(y_{f}|k_x)C_n(y_{f}|k_n)}{p(y_{f}|k_x,k_n)}
\label{e:spp_kx_kn}
\end{align}
Finally, using (\ref{e:mask-post-prob}) and (\ref{e:spp_kx_kn}), the localisation weight (\ref{e:mask-double-sum}) becomes  a weighted average of the \glspl{tpp} in (\ref{e:spp_kx_kn}) for all possible combinations of $(k_x,k_n)$
\begin{equation}
\omega_{f}= \sum_{k_x,k_n} \frac{\gamma^{(k_x,k_n)} p_x(y_{f}|k_x)C_n(y_{f}|k_n)}{p_x(y_{f}|k_x)C_n(y_{f}|k_n) + p_n(y_{f}|k_n)C_x(y_{f}|k_x)}.
\label{e:weights_final}
\end{equation}


\subsection{Adaptation to power level differences}
\label{ss:gmm_adaptation}

The mismatch in the power level of acoustic sources between training and testing conditions may cause the source \gls{gmm} to inaccurately represent the spectral characteristics of the sources in the testing condition. To alleviate this problem, we introduce an algorithm that adapts the source \glspl{gmm} directly from the noisy signals on-the-fly, before estimating the localisation weights from the same signal. To simplify the discussion, here we only consider adaptation of the \gls{gmm} for the interfering sources (i.e. we assume that the power level of the target source does not change). However, the extension of the procedure to also adapt the target source \gls{gmm} is straightforward. 

We assume that all the mixture compoments in the \gls{gmm} are adapted by the same level as follows:
\begin{equation}
p(\vec{y}|\lambda_n)= \sum_{k} P(k|\lambda_n) \mathcal{N}\left(\vec{y}; \vec{\mu}_n^{(k)} + \vec{\beta}, \mat{\Sigma}_n^{(k)}\right)
\end{equation}
where $\vec{\beta}= [\beta_{1}, \dots, \beta_{D}]$ is the vector that accounts for the power level differences in each frequency channel. Note that only the means are adapted and variances are kept the same. To determine $\vec{\beta}$ directly from the noisy observations, we resort to an iterative approach based on the \gls{em} algorithm \cite{Dem1977}. Let $\hat{\vec{\beta}}$ denote the current estimate of \vec{\beta}. The function to be optimised is
\begin{align}
Q(\vec{\beta}, \hat{\vec{\beta}})= \sum_t \sum_{k_x} \sum_{k_n} \gamma_t^{(k_x,k_n)} \sum_f \log p(y_{tf}|k_x,k_n,\beta_f)
\label{e:em-q-function}
\end{align}
where $\gamma_t^{(k_x,k_n)} $ is the posterior probabillity in (\ref{e:mask-post-prob}) and is computed using the current estimate $\hat{\vec{\beta}}$. $p(y_{tf}|k_x,k_n,\beta_f)$ is given by (\ref{e:prob_obs_y}).

By setting the derivative $\partial Q(\vec{\beta}, \hat{\vec{\beta}})/\partial \beta_f$ equal to zero, we obtain the following updating equation for $\beta_f$:
\begin{align}
\beta_f = \frac{1}{T} \sum\limits_t \sum\limits_{k_x} \sum\limits_{k_n} \gamma_t^{(k_x,k_n)}  \left( \overline{n}_{tf}^{(k_x,k_n)}-\mu_{nf}^{(k_n)} \right)
\label{e:beta-update}
\end{align}
where $\overline{n}_{tf}^{(k_x,k_n)}$ is the estimate of $n_{tf}$,
\begin{align}
\overline{n}_{tf}^{(k_x,k_n)} &= \alpha_{tf}^{(k_x,k_n)} \tilde{n}_{tf}^{(k_n)} +  \left(1- \alpha_{tf}^{(k_x,k_n)}\right) y_{tf}.
\end{align}

In the last equation we use the short-hand notation $\alpha_{tf}^{(k_x,k_n)} \triangleq  P(x_{f}=y_{f}, n_{f} \le y_{f}|k_x,k_n,\hat{\beta}_f, \vec{y})$ for the \gls{tpp} in 
(\ref{e:spp_kx_kn}). $\tilde{n}_{tf}^{(k_n)}$ represents the estimate for $n_{tf}$ under the assumption that $n_{tf}$ is masked by the target source energy $x_{tf}$. In this case, $n_{tf}$ is upper-bounded by the observation $y_{tf}$. Thus, $\tilde{n}_{tf}^{(k_n)}$ corresponds to the following expected value,
\begin{equation}
\tilde{n}_{tf}^{(k_n)} = \int_{-\infty}^{y_{tf}} n_{tf} \, p(n_{tf}|k_n,\hat{\beta}_f) \, dn_{tf}.
\end{equation}

From (\ref{e:beta-update}), it can be seen that the updated value for $\beta_f$ is a weighted average of the power level differences between the estimates $\overline{n}_{tf}^{(k_x,k_n)}$ and the \gls{gmm} means $\mu_{nf}^{(k_n)}$. This equation is iteratively applied until $\beta_f$ converges.

%% file: 04_Evaluation.tex

\section{Evaluation}
\label{s:eval}

To evaluate the proposed binaural localisation system, a virtual acoustic environment was created which contained a target speech source and a masking source selected from a number of different noise types.

\subsection{Binaural simulations}

Binaural audio signals were created by convolving monaural signals with \glspl{hrir} or \glspl{brir}. An \gls{hrir} catalog based on the \gls{kemar} dummy head~\cite{WierstorfGeierRaakeSpors11} was used for simulating the anechoic training signals. The evaluation stage used the Surrey \gls{brir} database~\cite{HummersoneMasonBrookes10} to simulate reverberant room conditions. The Surrey database was captured using a Cortex \gls{hats} and includes four room conditions with various amounts of reverberation (see Table~\ref{t:rooms} for the reverberation time (T$_{60}$) and the \gls{drr} of each room). Binaural mixtures of two simultaneous sources were created by convolving each source signal with \glspl{brir} separately before adding them together in each of the two binaural channels.

\begin{table}[thb]
\caption{Room characteristics of the Surrey \gls{brir} database~\cite{HummersoneMasonBrookes10}.}
\label{t:rooms}
\vspace{1mm}
\centering
\begin{tabular}{@{} l  c  c  c  c  @{}}
\hline\hline
 & Room A & Room B  & Room C & Room D \\
\hline
$\mathrm{T}_{60}$ (s) & 0.32 & 0.47 & 0.68 & 0.89  \\
$\mathrm{DRR}$ ($\deci\bel$) & 6.09 & 5.31 & 8.82 & 6.12 \\
\hline\hline
\end{tabular}
\end{table}

\subsection{Target and masker signals}

The target source signals were drawn from the GRID speech corpus~\cite{GridCorpus}. Each GRID sentence is approximately $2$ sec long and has a six-word form (e.g., ``\textit{lay red at G 9 now}''). Both a male talker and a female talker were selected as the target speech source. All target signals were sampled at 16\,kHz.

\begin{figure}[t]
\centerline{\includegraphics[trim=3.8cm 2.1cm 3.9cm 1.9cm, clip=true,width=\columnwidth]{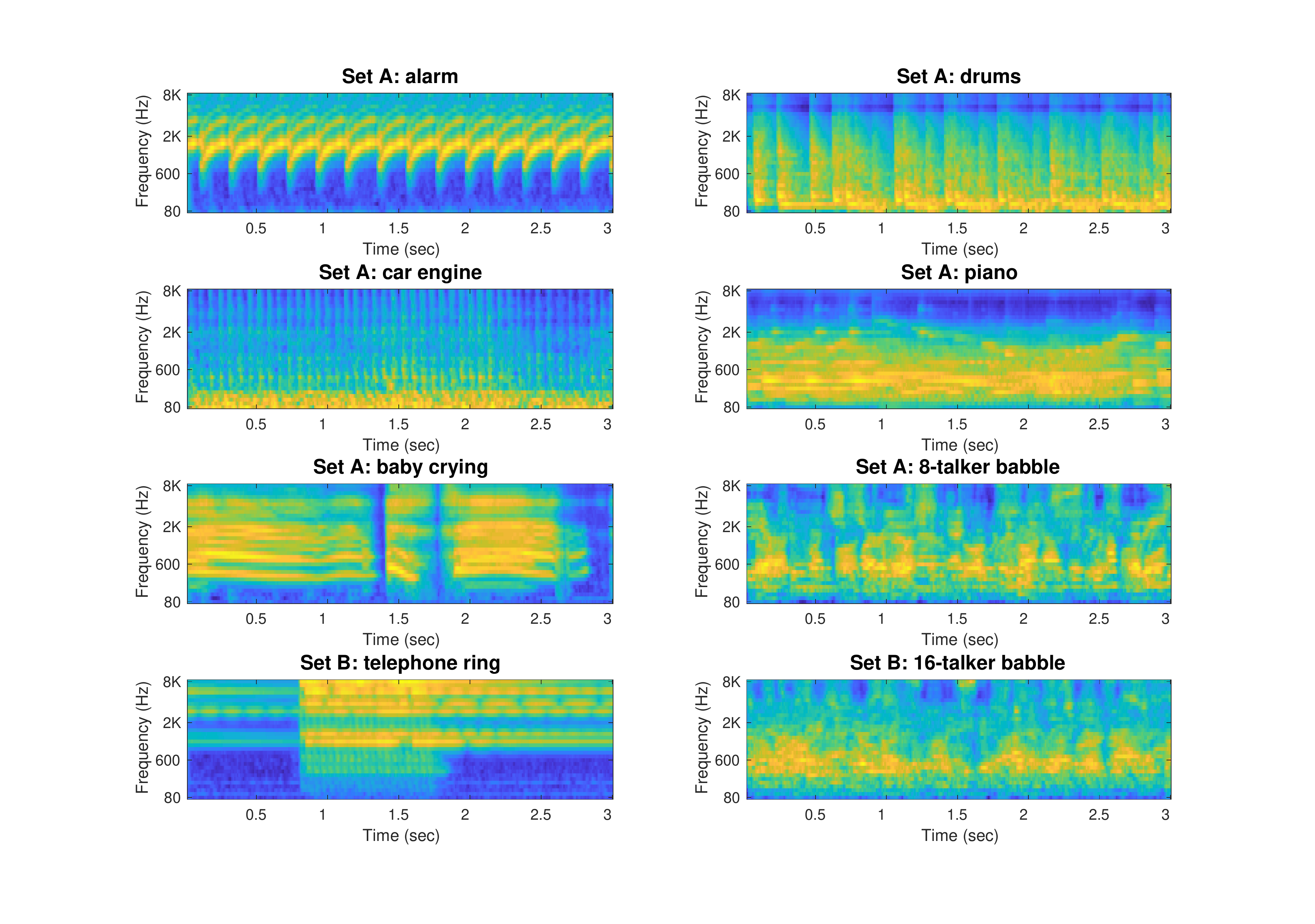}}
\caption{Ratemap representations of various masker sounds used in this study.}
\label{f:noise}
\end{figure}

In our previous study \cite{MaEtAl2015topdown}, all noise types were used to estimate the \gls{ubm}. In order to investigate how well the system is able to deal with unseen noise types, two sets of noise signals were used as the masker source. Noise Set A consists of six masker types with various degrees of spectro-temporal complexities, and were used to estimate the \gls{ubm}. Noise Set B consists of two additional masker types as the ``unseen noise'' set, i.e. they were excluded for training the \gls{ubm}. The details of both noise sets are summarised in Tables~\ref{t:noise_set_A} and \ref{t:noise_set_B}. In all cases, noise signals were 90 seconds long and were sampled at 16\,kHz. Fig.~\ref{f:noise} shows example ratemap representations of these noise signals.

\begin{table}[thb]
\def\arraystretch{1.3}
\caption{Descriptions of masker sounds used in Noise Set A.} 
\label{t:noise_set_A}
\vspace{1mm}
\centering
\begin{tabular}{ p{1.1cm}  p{6.9cm} }
\hline\hline
alarm & car alarm sound, rhythmic moderate-narrow-band signal between 600\,Hz and 3\,kHz.\\
\hline
drums & drumming sound, strong onsets synchronised across frequency, and with significant energy distributed between 100--300~Hz.\\
\hline
car engine & highly modulated at 15~Hz with most energy distributed below 300~Hz.\\
\hline
piano & fast playing solo piano sound, with most energy distributed below 2~KHz.\\
\hline
baby crying & crying baby sound, high pitch, less rhythmic with harmonics often lasting longer than 1 second.\\
\hline
16-talker babble & created from 16 talkers randomly selected from the TIMIT corpus~\cite{Gar1993}, mostly overlapping, speech frequency range.\\
\hline\hline
\end{tabular}
\end{table}  

\begin{table}[thb]
\def\arraystretch{1.3}
\caption{Descriptions of masker sounds used in Noise Set B.} 
\label{t:noise_set_B}
\vspace{1mm}
\centering
\begin{tabular}{ p{1.1cm}  p{6.9cm} }
\hline\hline
telephone ring &  telephone ring, periodic and narrow-band sound with significant energy at around 1\,kHz and above 3\,kHz.\\
\hline
32-talker babble & created from 32 talkers randomly selected from the TIMIT corpus, mostly overlapping, speech frequency range.\\
\hline\hline
\end{tabular}
\end{table}  

\subsection{Localisation DNN training}

As shown in previous studies~\cite{WoodruffWang2012,MayvandeParKohlrausch13,MayMaBrown2015}, \gls{mct} can increase the robustness of localisation systems in reverberant multi-source conditions. To train the \glspl{dnn}, in this study a \gls{mct} dataset was created by mixing the training signals at a specified azimuth with diffuse noise as described in \cite{MayMaBrown2015}. The diffuse noise consisted of 72 uncorrelated, white Gaussian noise sources that were placed across the full 360$\deg$ azimuth range in steps of $5\deg$. Both the target signals and the diffuse noise were created by using the same anechoic \gls{hrir} recorded using a \gls{kemar} dummy head~\cite{WierstorfGeierRaakeSpors11}.

Following \cite{MaEtAl2017dnn}, the localisation training data consisted of speech sentences from the {TIMIT} database~\cite{Gar1993}. A set of 30 speech sentences was randomly selected for each of the 72 azimuth locations, equally distributed in the 360$\deg$ range. For each spatialised training sentence, the anechoic signal was corrupted with diffuse noise at three \glspl{snr} ($20$, $10$ and $0\,\deci\bel$). The corresponding binaural features (\glspl{itd}, \gls{ccf}, and \glspl{ild}) were then extracted. Only those features for which the \textit{a priori} \gls{snr} between the target and the diffuse noise exceeded $-5\,\deci\bel$ were used for training. This negative \gls{snr} criterion ensured that the multi-modal clusters in the binaural feature space at higher frequencies, which are caused by periodic ambiguities in the cross-correlation function, were properly captured. The localisation DNNs were not retrained for the target source drawn from the GRID speech corpus.

\subsection{Source model training}
\label{ss:source_model_training}

This study is concerned with binaural sound localisation. In order to learn a source spectral model in a binaural setting, binaural source signals were created by convolving monaural signals with the Room A \gls{brir} from the Surrey database~\cite{HummersoneMasonBrookes10} for azimuths ranging between $\left[-90\deg, 90\deg\right]$ in $5\deg$ steps. Ratemap features were first extracted from each of the binaural channels and then averaged across the two channels. The ratemap features for all the azimuths were used to train source models using the EM algorithm as described in Section~\ref{ss:source_model}.

The identity of the attended target source is assumed known \textit{a priori} in this study. For the target speech source, 90 seconds of training data were used to train a 16-mixture target GMM.

For each noise source in Noise Set A, 4/5 of the 90-second signal was used as the training data and the remaining 1/5 was used as the test data. When the identity of the masker source is unavailable, the system uses a universal background model and performs adaptation on-the-fly during testing in order to better match the spectral profile of the masker source in a mixed test signal. All the training data from Noise Set A were used to train a 16-mixture \gls{ubm}. Noises in Set B were excluded in this process.

If the identity of the masker source is also available \textit{a priori}, the system can directly exploit the corresponding masker model. The number of mixtures for each GMM was selected based on its spectro-temporal complexity and is listed in Table~\ref{t:gmm_mix}.

\begin{table}[thb]
\def\arraystretch{1.3}
\caption{The number of Gaussian mixture components used for each source model.} 
\label{t:gmm_mix}
\vspace{1mm}
\centering
\begin{tabular}{@{} c c c c c c c c @{}}
\hline\hline
Target & \multicolumn{6}{c}{Noise Set A} & \\
\cline{2-7}
speech & alarm & drum & car & piano & baby & 8-talker & \multirow{-2}{*}{UBM} \\
\hline
16 & 2 & 2 & 2 & 3 & 3 & 4 & 8\\
\hline\hline
\end{tabular}
\end{table}  

\subsection{Experimental setup}

\begin{figure}[t!]
\center
\includegraphics[width=.7\columnwidth]{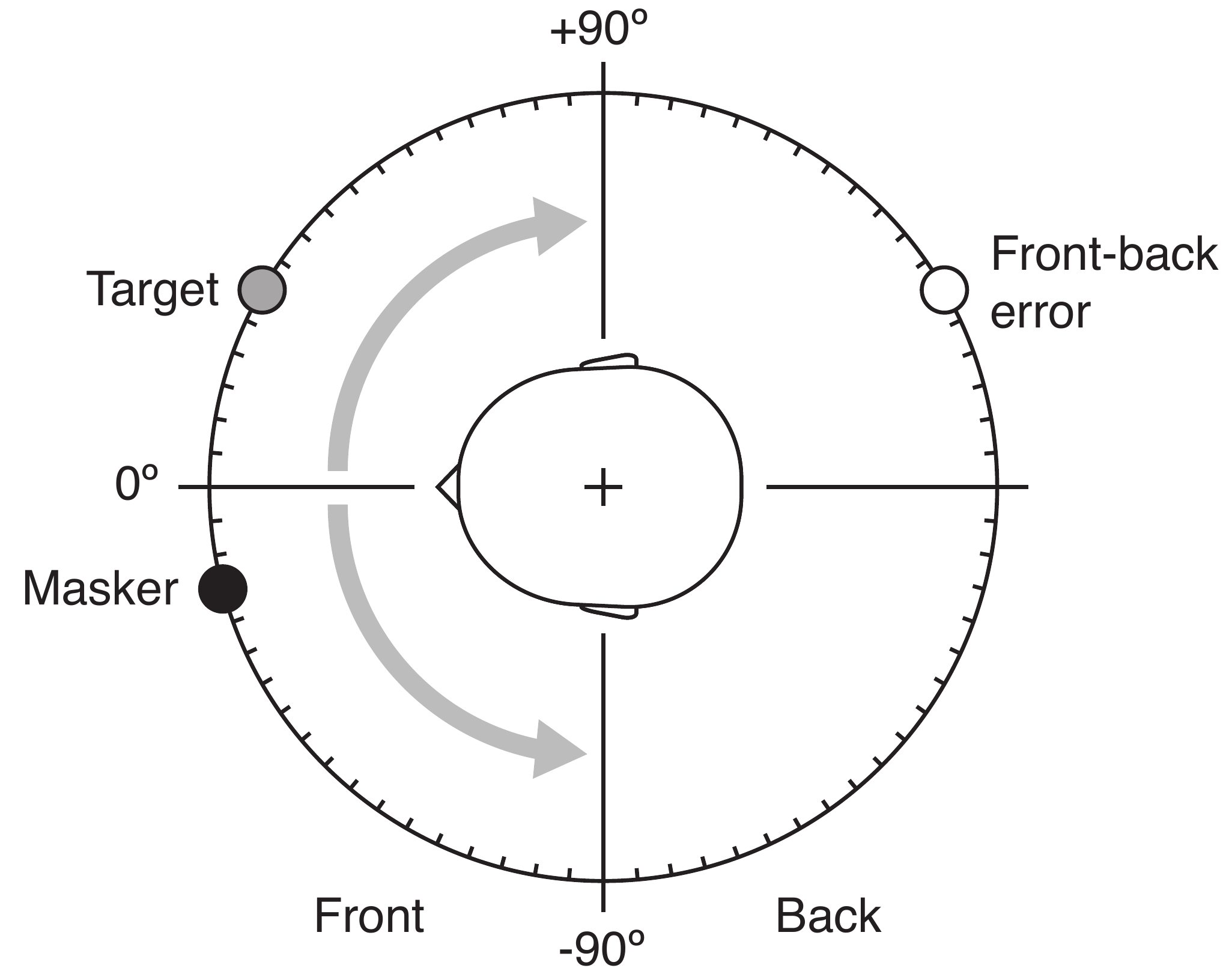}
\caption{Schematic diagram of the virtual listener configuration, showing a typical arrangement of target source (here at +30$\deg$) and masker (at -15$\deg$). Target source positions were limited to the range [-90$\deg$,+90$\deg$] as indicated by the gray arrows. Potentially, a target source in front of the head could be incorrectly attributed to a location behind the head -- a \textit{front-back error} -- as shown by the open circle.}
\label{f:azimuth}
\end{figure}

The evaluation set contained $50$ GRID sentences from the target speech source which were not included in training. For each target signal, a masker signal that matched the length of the target was randomly selected from the test set. The target source was then mixed with the masker signal in a binaural setting. As shown in Fig.~\ref{f:azimuth}, the target source varied in azimuth within the range of $\left[-90\deg, 90\deg\right]$ in $10\deg$ steps. However, this knowledge was not available to the systems and the full 360$\deg$ azimuth range could be reported (and hence front-back errors could occur, as shown in Fig.~\ref{f:azimuth}).

The azimuth of the masker was randomly selected each time from the same azimuth range, while ensuring an angular distance of at least $10\deg$ between the two competing sources. Source locations were limited to this azimuth range because the Surrey database only includes azimuths in the frontal hemifield. However, our localisation system was not provided in any case with information that the azimuth of the source lay within this range; it was free to report the azimuth within the full $360\deg$ range. Both target and masker signals were normalised by their \gls{rms} amplitudes prior to spatialisation at two target-to-masker ratios (TMRs): 0\,dB and -6\,dB.
 
The baseline system was a state-of-the-art binaural localisation system using DNNs~\cite{MaEtAl2017dnn}, with no model-based knowledge about the sources. The proposed localisation system exploiting model-based information employed the same localisation DNNs, but was given prior knowledge of the target source so that it could be ``attended to'' for more accurate localisation. The system was evaluated in three scenarios:

\subsubsection{The identity of the masker source was assumed to be available \textit{a priori}}
The corresponding masker source \gls{gmm} was used as the background model in order to estimate the set of localisation weights $\omega_{tf}$ in Eq.~(\ref{e:post-angle}). Noise Set A was used for evaluating this scenario.

\subsubsection{The masker source identity was unavailable but the source was used for training the \gls{ubm}}
In the second scenario, the \gls{ubm} estimated from all the noise types in Noise Set A was used as the background model.

\subsubsection{The masker source identity was unavailable and the source was not used for training the \gls{ubm}}
Similar to scenario 2, in this scenario the \gls{ubm} estimated from all the noise types in Noise Set A was used as the background model. However, to test how well the system could generalise unseen noise types, Noise Set B was used for evaluation, which was not used for estimating the \gls{ubm}.

In all scenarios, the background model (either the correct masker model or the \gls{ubm}) was employed with or without adaptation.


For all the evaluated systems, the number of competing sources (two in this study) was assumed to be known \emph{a priori}. Therefore each system reported two estimated source azimuths. The localisation performance was measured for the target source only by comparing true target source azimuths with the estimated azimuths. The \emph{target localisation error rate} was measured by counting the number of sentences for which one of the azimuth estimates was outside a predefined grace boundary of $\pm 5\deg$ with respect to the true target azimuths.

%% file: 05_Results.tex

\section{Results and Discussion}
\label{s:results}

\begin{figure}[thb]
\centerline{\includegraphics[trim=3cm 3.2cm 3.3cm 2.1cm, clip=true,width=\columnwidth]{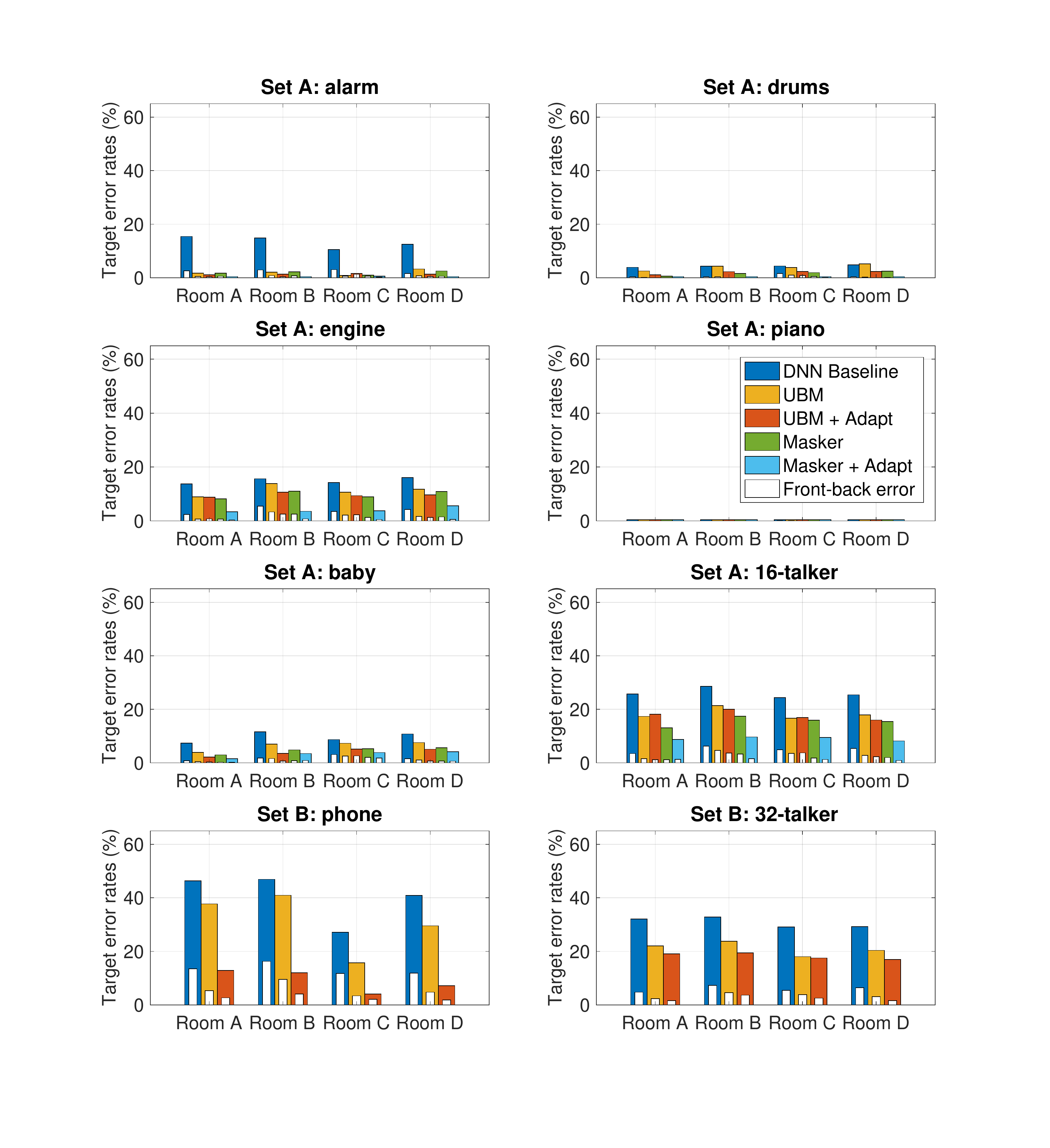}}
\caption{Localisation error rates for localising the target source in the presence of various maskers, at a target-to-masker ratio (TMR) of 0\,dB.  The proportions of front-back errors are indicated as white bars.}
\label{f:loc_err_0dB}
\end{figure}

\begin{figure}[thb]
\centerline{\includegraphics[trim=3cm 3.2cm 3.3cm 2.1cm, clip=true,width=\columnwidth]{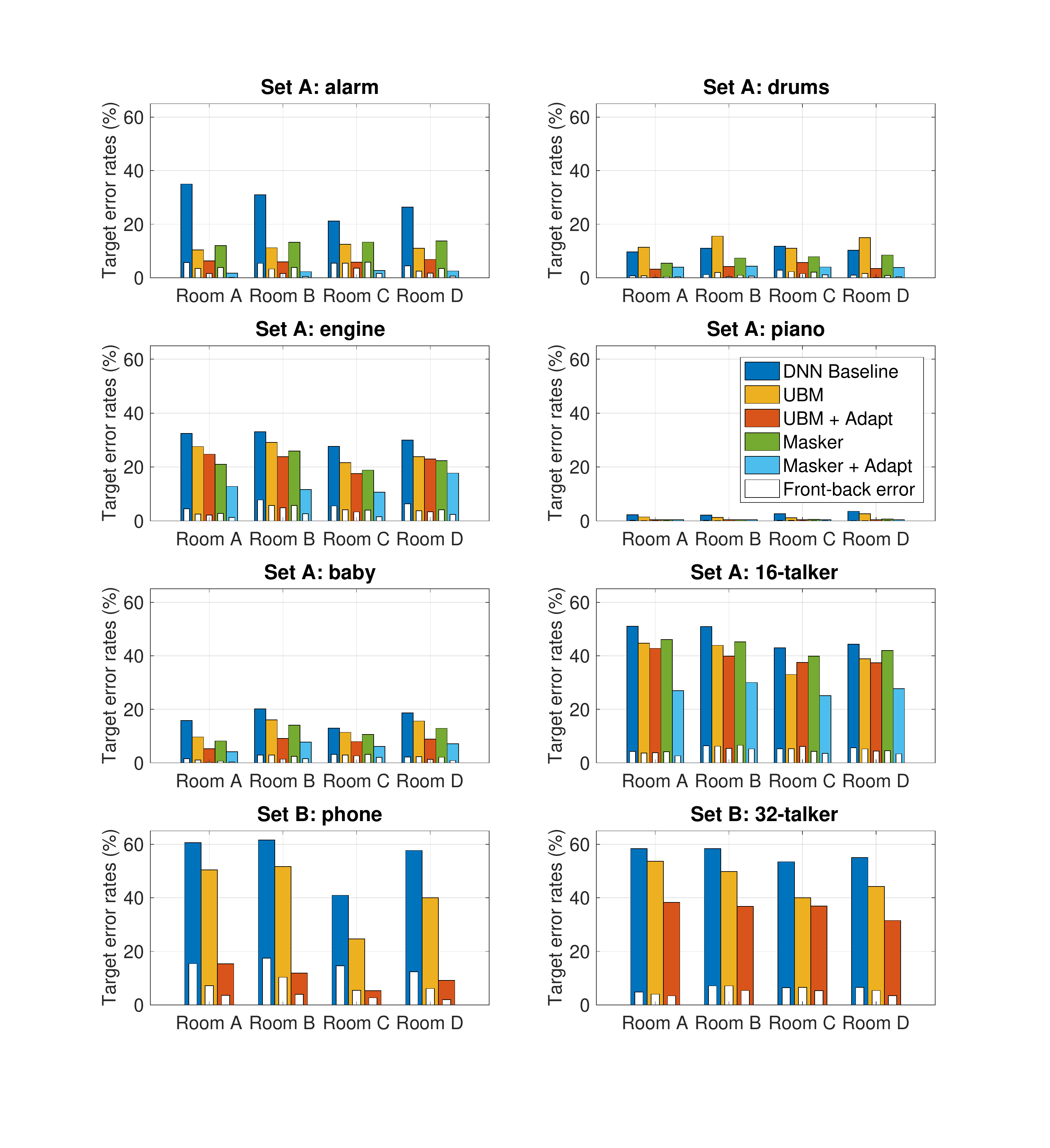}}
\caption{Localisation error rates for localising the target source in the presence of various maskers, at a target-to-masker ratio (TMR) of -6\,dB. The proportions of front-back errors are indicated as white bars.}
\label{f:loc_err_-6dB}
\end{figure}

The target localisation error rates of various binaural localisation systems are shown in Figs.~\ref{f:loc_err_0dB} and \ref{f:loc_err_-6dB} for the 0\,dB and the -6\,dB TMR cases, respectively. The results are averaged between the male speech source case and the female speech source case.

For each of the known masker types (top 6 maskers), the results using both UBM and the respective masker model were shown while for the unknown masker types in Set B only the results using the UBM are shown. Furthermore, since in this study the system did not assume that the sound sources are located in the frontal hemifield, the front-back error rates are shown as white bars in each figure.

The performance of the baseline system, which uses no model-based information for localisation, varied greatly across different masker conditions. The results show that the baseline system performed worse in masker conditions such as `phone', `alarm', and N-talker babble. While it is straightforward to understand why localisation of the target speech is less reliable in N-talker babble, as the masker spectrum overlaps that of the target speech, the poor performance in the `phone' and `alarm' conditions requires further explanation. This is likely due to two reasons. Firstly, these sounds are narrowband as shown in Fig.~\ref{f:noise}, and the target speech was completely masked at the frequencies where the maskers's energy dominated. Hence, in these frequency bands only small glimpses were available to localise the target source and the baseline system tended to report high probabilities for the masker azimuths. Secondly, most of the energy for the narrowband sounds is located at high frequencies. The baseline system employed cross-correlation features and ILDs as localisation features, and these features are known to be more reliable in high frequencies~\cite{MaEtAl2017dnn}. In particular, the ILDs are more pronounced at high frequencies due to the size of the head compared to the wavelength of incoming sounds~\cite{Blauert97}. When these frequency bands were dominated by the masker, the system was more likely to report the location of the masker, especially when the global TMR was negative. 
Detailed analysis of the results shows that most localisation errors were due to incorrectly reporting the masker locations.

In contrast, the target localisation error rates were lower when the masker source dominates low frequencies. For example, in the `piano' and `drums' conditions, the target localisation error rates were below 10\% at the TMR of 0\,dB, and even at the TMR of -6\,dB, the localisation error rates were still below 5\% in the `piano' condition. This is largely because the maskers dominated frequency regions that were less reliable for localisation with the DNN system, and thus the performance remained robust in the presence of the masker.

When model-based source knowledge was employed (in Figs.~\ref{f:loc_err_0dB} and \ref{f:loc_err_-6dB}, UBM or Masker), the localisation system was able to identify the spectro-temporal regions dominated by the target speech. Therefore the system was more likely to report the location of the target source by weighing those regions more. Figs.~\ref{f:loc_err_0dB} and \ref{f:loc_err_-6dB} show that the results significantly improve when model-based information was used for localisation, particularly for narrowband sounds such as `alarm'. 

\subsection{Effect of using the masker model vs. the UBM}


For each masker source in Noise Set A, a corresponding masker model was created. If the masker type is assumed to be known, the correct masker model was used as the background model, otherwise the UBM was used. Across different masker conditions, the use of a background model greatly improved the target localisation accuracy. Comparing the results in Fig.~\ref{f:loc_err_0dB} using the UBM and the correct masker model, without adaptation, one can see that at 0\,dB TMR using the UBM the system performed comparably with that using the correct masker model. This is largely because the UBM was trained using signals of the masker types from Noise Set A. The use of the \gls{gmm} was effective to capture the spectral profiles of all the maskers and this helped localise the target source using the proposed framework. 

As can be seen in Fig.~\ref{f:loc_err_-6dB}, at -6\,dB TMR, the localisation error reduction using the correct masker model becomes larger. This is expected, as a more detailed masker model will produce more reliable localisation weights than the general \gls{ubm}. However, the use of the \gls{ubm} minimises the assumptions made about the active masker sources. Such a system is potentially more suitable for an attention-driven model of sound localisation, in which the attended target source may be switched and the localisation weights can be dynamically recomputed in order to localise the newly attended source.

\subsection{Effect of unseen masker types}

In Noise Set A, when the identity of the masker was assumed unknown, the \gls{ubm} was used as the background model. However, the parameters of the \gls{ubm} were estimated using all the masker types from Noise Set A. To evaluate how well the system could generalise to unseen masker types, Noise Set B was excluded from training the \gls{ubm}. Comparing the \gls{ubm} results for Noise Sets A and B, it is clear that using a \gls{ubm} the system did not improve the target localisation accuracy for Noise Set B as much as for Noise Set A. This is especially the case for narrowband maskers.  In the unseen `phone' condition, the average target localisation error rate was reduced from 41\% to 31\% at the TMR of 0\,dB, and   was reduced from 55\% to 42\% at the TMR of -6\,dB. In contrast, in the `alarm' condition, the localisation error rate was reduced from 14\% to 2\% at 0\,dB TMR and from 30\% to 12\% at -6\,dB TMR. The error reduction was more similar between the 16-talker and 32-talker conditions, probably because the two speech babbles are more similar in spectral shapes. This suggests that the \gls{ubm} worked less effectively for masker conditions that were not used for training the \gls{ubm}, but it is still beneficial to use the \gls{ubm} in unseen masker conditions.

\subsection{Effect of background-model adaptation}

As demonstrated by the results, background-model adaptation had a major contribution in reducing the target localisation error rates across various conditions. At the TMR of 0\,dB the energy level mismatch between the trained background models and the test signals was minimal, but the adaptation process was still beneficial for Noise Set A as it can accommodate the level difference of individual test signals. The improvement was larger when the background model was the correct masker model (as opposed to the \gls{ubm}), particularly for masker types that are difficult to model, such as `engine' (many unpredictable events) and `16-talker babble' (less stationary). For Noise Set B, the benefit of adaptation is more apparent, especially for the narrowband `phone' condition. This suggests that the adaptation process was also able to adapt the model parameters to match better the spectral profile of an unseen masker.

Background-model adaptation benefitted the systems more at the TMR of -6\,dB where there is a mismatch between the model level and the test signal level. Similarly to the 0\,dB TMR case, the improvement appeared larger when the correct masker model was used as the background model. This is likely because with the correct masker model only the energy level needs to be adapted, whereas with the \gls{ubm} the spectral profile also needs to be adapted. As the model parameters were adapted on-the-fly using a short test signal (less than 2-sec long) that was mixed with a masker sound, it is more difficult to reestimate the model parameters correctly.

\begin{figure}[thb]
\centerline{\includegraphics[trim=2cm 1.7cm 2.5cm 2cm, clip=true,width=\columnwidth]{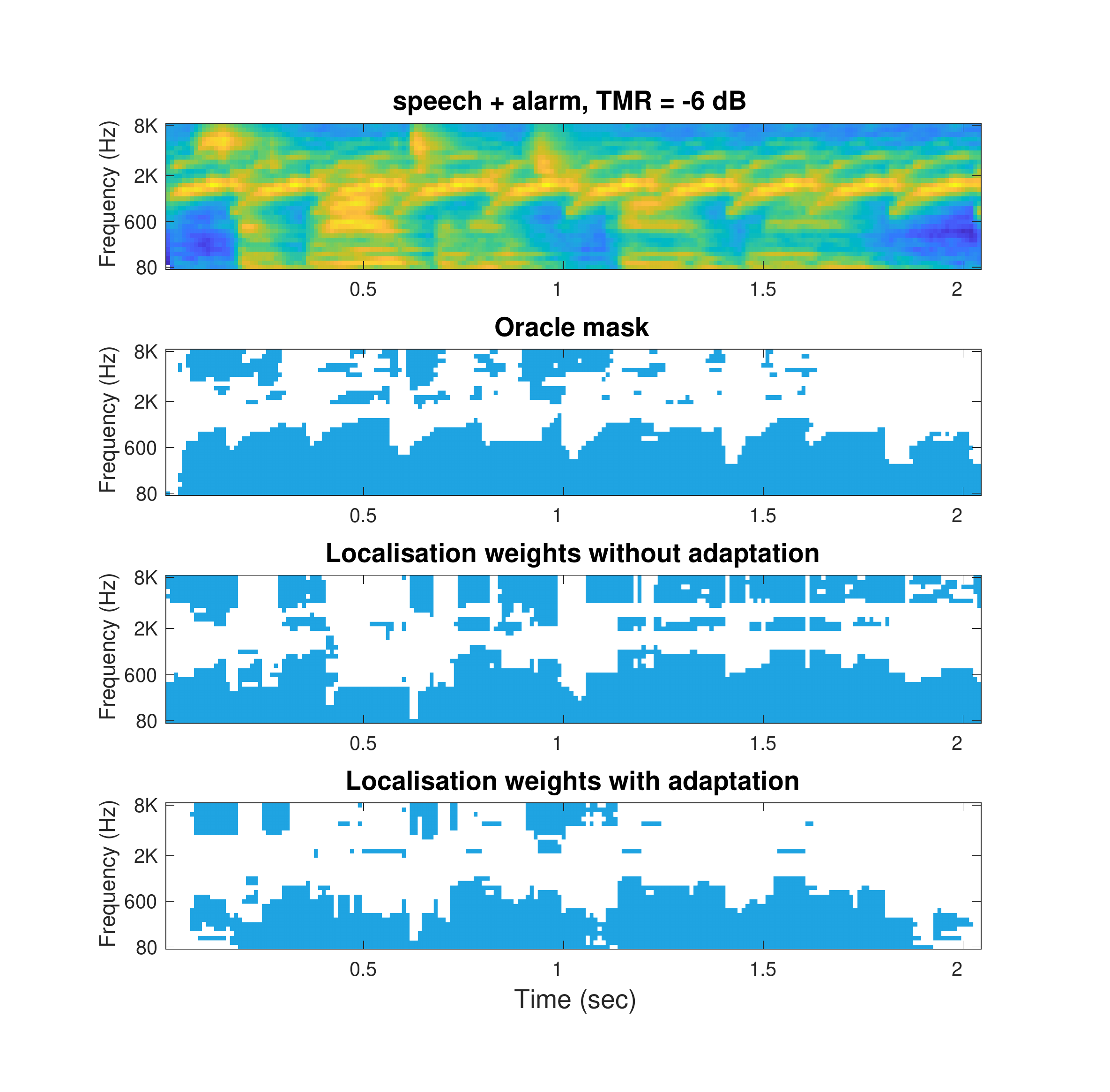}}
\caption{Localisation weights estimated for target speech mixed with alarm sound at a target-to-masker ratio (TMR) of -6\,dB using the UBM with and without adaptation. The `oracle' mask was shown to indicate the spectro-temporal regions dominated by the target speech (blue regions). The localisation weights larger than 0.5 were shown in blue in the bottom 2 panels.}
\label{f:loc_weights}
\end{figure}

To illustrate the effect of various stages in the proposed system, Fig.~\ref{f:loc_weights} shows an example of the estimated localisation weights. Here the target speech was mixed with an alarm sound at a TMR of -6\,dB. The \gls{ubm} was used when estimating the localisation weights. The `oracle' mask shows the spectro-temporal regions dominated by the target speech using \textit{a priori} information of the pre-mixed signals. The localisation weights estimated without adaptation bear some resemblance to the oracle mask, but incorrectly give more weight to the high frequency regions above 2\,kHz that are dominated by the masker. With adaptation these masker-dominated regions are given less weight, and the estimated mask is closer to the oracle mask.

%% file: 06_Conclusion.tex

\section{Conclusions}
\label{s:conc}

This paper has presented a novel computational framework for binaural sound localisation that combines data-driven and model-based information flow. By jointly exploiting model-based knowledge about the source spectral characteristics in the acoustic scene, the system is able to selectively weight binaural cues in order to more reliably localise the attended source. To address the mismatch between the training and testing conditions, a model adaptation process is employed on-the-fly which re-estimates the background model parameters directly from the noisy observations in an iterative manner. Evaluation using masker sources with varying spectro-temporal complexity, including masker types that are not seen during the training stage, showed that by exploiting source models in this way, sound localisation performance can be improved substantially under conditions where multiple sources and room reverberation are present.

One of the advantages of the proposed system is that information fusion across frequency bands is done after the DNNs estimate the azimuth posteriors. The late fusion of information allows use of single-source data for training, as otherwise the amount of required training data would increase exponentially with the number of sources. In this way, the framework is also able to generalise to localisation of other sounds and is not tailored to the type of sound used for training.

The current evaluation involved only two sound sources (a target source and an interfering source). In general, the proposed approach is able to generalise to the scenario where there are more than one interfering source. In this case, one would consider all interfering sources as a single masker and use the universal background model to model the combination of all interfering sources. The complexity of the framework stays the same in this case.

The proposed sound localisation framework could be combined with source identification in order to estimate the identity of the target source that the system `attends' to. Such an attention-driven model could be used to localise an attended source whose identity is not available \textit{a priori}, e.g. a talker that speaks a keyword in an acoustic mixture. Source localisation and source identification could then interact in an ongoing iterative process. In addition, the framework described here could be integrated with an approach that uses source models to `perceptually restore' parts of the target sound that have been masked \cite{remes2015}. Another future direction is the extension to cross-modal control. The proposed system is a general framework through which other modalities could also be incorporated, such as a vision system on a mobile robot.

%% file: paper.bbl
\begin{thebibliography}{10}
\providecommand{\url}[1]{#1}
\csname url@samestyle\endcsname
\providecommand{\newblock}{\relax}
\providecommand{\bibinfo}[2]{#2}
\providecommand{\BIBentrySTDinterwordspacing}{\spaceskip=0pt\relax}
\providecommand{\BIBentryALTinterwordstretchfactor}{4}
\providecommand{\BIBentryALTinterwordspacing}{\spaceskip=\fontdimen2\font plus
\BIBentryALTinterwordstretchfactor\fontdimen3\font minus
  \fontdimen4\font\relax}
\providecommand{\BIBforeignlanguage}[2]{{%
\expandafter\ifx\csname l@#1\endcsname\relax
\typeout{** WARNING: IEEEtran.bst: No hyphenation pattern has been}%
\typeout{** loaded for the language `#1'. Using the pattern for}%
\typeout{** the default language instead.}%
\else
\language=\csname l@#1\endcsname
\fi
#2}}
\providecommand{\BIBdecl}{\relax}
\BIBdecl

\bibitem{Bre1990}
A.~Bregman, \emph{Auditory Scene Analysis}.\hskip 1em plus 0.5em minus
  0.4em\relax Cambridge, MA: MIT Press, 1990.

\bibitem{WangBrown2006}
D.~L. Wang and G.~J. Brown, Eds., \emph{Computational Auditory Scene Analysis:
  Principles, Algorithms and Applications}.\hskip 1em plus 0.5em minus
  0.4em\relax Wiley/IEEE Press, 2006.

\bibitem{spence1994}
G.~C. Spence and J.~Driver, ``Covert spatial orienting in audition: exogenous
  and endogenous mechanisms,'' \emph{Journal of Experimental Psychology},
  vol.~20, pp. 555--574, 1994.

\bibitem{Bronkhorst2015}
A.~Bronkhorst, ``The cocktail-party problem revisited: early processing and
  selection of multi-talker speech,'' \emph{Attention, Perception, and
  Psychophysics}, vol.~77, no.~5, pp. 1465--1487, 2015.

\bibitem{gaese2002}
B.~H. Gaese and H.~Wagner, ``Precognitive and cognitive elements in sound
  localization,'' \emph{Zoology}, vol. 105, pp. 329--339, 2002.

\bibitem{Winkowski:2006rm}
D.~E. Winkowski and E.~I. Knudsen, ``Top-down gain control of the auditory
  space map by gaze control circuitry in the barn owl.'' \emph{Nature}, vol.
  439, no. 7074, pp. 336--339, 2006.

\bibitem{MaEtAl2017dnn}
N.~Ma, T.~May, and G.~J. Brown, ``Exploiting deep neural networks and head
  movements for robust binaural localisation of multiple sources in reverberant
  environments,'' \emph{{IEEE} Trans. Audio, Speech, Lang. Process.}, vol.~25,
  no.~12, pp. 2444--2453, 2017.

\bibitem{Chr2009}
H.~Christensen, N.~Ma, S.~Wrigley, and J.~Barker, ``A speech fragment approach
  to localising multiple speakers in reverberant environments,'' in \emph{Proc.
  IEEE Int. Conf. Acoust., Speech, Signal Process.}, Taipei, 2009, pp.
  4593--4596.

\bibitem{MandelEtAl2010}
M.~Mandel, R.~Weiss, and D.~Ellis, ``Model-based expectation maximization
  source separation and localization,'' \emph{{IEEE} Trans. Audio, Speech,
  Lang. Process.}, vol.~18, no.~2, pp. 382--394, 2010.

\bibitem{WoodruffWang2012}
J.~Woodruff and D.~L. Wang, ``Binaural localization of multiple sources in
  reverberant and noisy environments,'' \emph{{IEEE} Trans. Audio, Speech,
  Lang. Process.}, vol.~20, no.~5, pp. 1503--1512, 2012.

\bibitem{WoodruffWang2013}
J.~Woodruff and D.~Wang, ``Binaural detection, localization, and segregation in
  reverberant environments based on joint pitch and azimuth cues,'' \emph{IEEE
  T. Audio. Speech.}, vol.~21, no.~4, pp. 806--815, 2013.

\bibitem{Bar2005}
J.~Barker, M.~Cooke, and D.~Ellis, ``Decoding speech in the presence of other
  sources,'' \emph{Speech Commun.}, vol.~45, pp. 5--25, 2005.

\bibitem{Ma2013}
N.~Ma, J.~Barker, H.~Christensen, and P.~Green, ``A hearing-inspired approach
  for distant-microphone speech recognition in the presence of multiple
  sources,'' \emph{Comput. Speech. Lang.}, vol.~27, no.~3, pp. 820--836, 2013.

\bibitem{liu2011}
J.~Liu, H.~Erwin, and G.-Z. Yang, ``Attention driven computational model of the
  auditory midbrain for sound localization in reverberant environments,'' in
  \emph{Neural Networks (IJCNN), The 2011 International Joint Conference on},
  July 2011, pp. 1251--1258.

\bibitem{MaEtAl2015topdown}
N.~Ma, G.~J. Brown, and J.~A. Gonzalez, ``Exploiting top-down source models to
  improve binaural localisation of multiple sources in reverberant
  environments.'' in \emph{Proc. Interspeech}, 2015, pp. 160--164.

\bibitem{MayvandeParKohlrausch11a}
T.~May, S.~van~de Par, and A.~Kohlrausch, ``A probabilistic model for robust
  localization based on a binaural auditory front-end,'' \emph{{IEEE} Trans.
  Audio, Speech, Lang. Process.}, vol.~19, no.~1, pp. 1--13, 2011.

\bibitem{Blauert97}
J.~Blauert, \emph{Spatial hearing - The psychophysics of human sound
  localization}.\hskip 1em plus 0.5em minus 0.4em\relax Cambride, MA, USA: The
  {MIT} Press, 1997.

\bibitem{MaEtAl2015dnn}
N.~Ma, G.~J. Brown, and T.~May, ``Robust localisation of of multiple speakers
  exploiting deep neural networks and head movements,'' in \emph{Proc.
  Interspeech}, 2015, pp. 3302--3306.

\bibitem{Bro1994}
G.~Brown and M.~Cooke, ``Computational auditory scene analysis,'' \emph{Comput.
  Speech. Lang.}, vol.~8, pp. 297--336, 1994.

\bibitem{WoodruffWang12}
J.~Woodruff and D.~L. Wang, ``Binaural localization of multiple sources in
  reverberant and noisy environments,'' \emph{{IEEE} Trans. Audio, Speech,
  Lang. Process.}, vol.~20, no.~5, pp. 1503--1512, 2012.

\bibitem{MayvandeParKohlrausch13}
T.~May, S.~van~de Par, and A.~Kohlrausch, ``Binaural localization and detection
  of speakers in complex acoustic scenes,'' in \emph{The technology of binaural
  listening}, J.~Blauert, Ed.\hskip 1em plus 0.5em minus 0.4em\relax
  Berlin--Heidelberg--New York NY: Springer, 2013, ch.~15, pp. 397--425.

\bibitem{Var1990}
A.~Varga and R.~Moore, ``Hidden {M}arkov model decomposition of speech and
  noise,'' in \emph{Proc. IEEE Int. Conf. Acoust., Speech, Signal Process.},
  1990, pp. 845--848.

\bibitem{Ren2010}
S.~J. Rennie, J.~R. Hershey, and P.~A. Olsen, ``Single-channel multitalker
  speech recognition,'' \emph{IEEE Signal Process. Mag.}, vol.~27, pp. 66--80,
  2010.

\bibitem{Reddy2004}
A.~M. Reddy and B.~Raj, ``Soft mask estimation for single channel speaker
  separation,'' in \emph{ISCA Tutorial and Research Workshop (ITRW) on
  Statistical and Perceptual Audio Processing}, 2004.

\bibitem{Reddy2007}
------, ``Soft mask methods for single-channel speaker separation,''
  \emph{{IEEE} Trans. Audio, Speech, Lang. Process.}, vol.~15, no.~6, pp.
  1766--1776, 2007.

\bibitem{Gonzalez2017}
J.~A. Gonzalez, A.~M. G{\'o}mez, A.~M. Peinado, N.~Ma, and J.~Barker,
  ``Spectral reconstruction and noise model estimation based on a basking model
  for noise robust speech recognition,'' \emph{Circ. Sys. Signal Pr.}, pp.
  1--30, 2017.

\bibitem{Dem1977}
A.~Dempster, N.~Laird, and D.~Rubin, ``Maximum likelihood from incomplete data
  via the em algorithm,'' \emph{Journal of the Royal Statistical Society.
  Series B}, vol.~39, no.~1, pp. 1--38, 1977.

\bibitem{WierstorfGeierRaakeSpors11}
H.~Wierstorf, M.~Geier, A.~Raake, and S.~Spors, ``A free database of
  head-related impulse response measurements in the horizontal plane with
  multiple distances,'' in \emph{Proc. 130th Conv. Audio Eng. Soc.}, 2011.

\bibitem{HummersoneMasonBrookes10}
C.~Hummersone, R.~Mason, and T.~Brookes, ``Dynamic precedence effect modeling
  for source separation in reverberant environments,'' \emph{{IEEE} Trans.
  Audio, Speech, Lang. Process.}, vol.~18, no.~7, pp. 1867--1871, 2010.

\bibitem{GridCorpus}
M.~Cooke, J.~Barker, S.~Cunningham, and X.~Shao, ``An audio-visual corpus for
  speech perception and automatic speech recognition,'' \emph{J. Acoust. Soc.
  Am.}, vol. 120, pp. 2421--2424, 2006.

\bibitem{Gar1993}
J.~S. Garofolo, L.~F. Lamel, W.~M. Fisher, J.~G. Fiscus, D.~S. Pallett, and
  N.~L. Dahlgren, ``{DARPA} {TIMIT} {A}coustic-phonetic continuous speech
  corpus {CD-ROM},'' \emph{National Inst. Standards and Technol. ({NIST})},
  1993.

\bibitem{MayMaBrown2015}
T.~May, N.~Ma, and G.~J. Brown, ``Robust localisation of multiple speakers
  exploiting head movements and multi-conditional training of binaural cues,''
  in \emph{Proc. ICASSP}, 2015, pp. 2679--2683.

\bibitem{remes2015}
U.~Remes, A.~L{\'o}pez, L.~Juvela, K.~Palom{\"a}ki, G.~J. Brown, P.~Alku, and
  M.~Kurimo, ``Comparing human and automatic speech recognition in a perceptual
  restoration experiment,'' \emph{Computer Speech and Language}, vol.~35, pp.
  14--31, June 2015.

\end{thebibliography}
